\documentclass[12pt]{article}
\usepackage{epsfig}
\renewcommand{\title}{        Holographic Cosmology and its Relevant Degrees
                                of Freedom
}
\advance\textheight by \topskip
\renewcommand{\baselinestretch}{1.2}
\renewcommand{\thefootnote}{\fnsymbol{footnote}}

\newcommand{\beq}{\begin{equation}}
\newcommand{\eeq}{\end{equation}}
\newcommand{\bea}{\begin{eqnarray}}
\newcommand{\eea}{\end{eqnarray}}

\begin{document}
\newtheorem{fig}[figure]{Figure}


\renewcommand{\baselinestretch}{1}
\renewcommand{\thefootnote}{\alph{footnote}}

\thispagestyle{empty}

\vspace*{-0.3cm} {\bf \hfill LBNL--43788}

\vspace*{-0.3cm} {\bf \hfill July 99} \vspace*{1.5cm}
{\Large\bf \begin{center} \title \end{center}}
{\begin{center}

\vspace*{0.5cm}
   {\begin{center} {\large\sc
                Richard Dawid\footnote{\makebox[1.cm]{Email:}
                                  dawid@thsrv.lbl.gov}}

    \end{center} } 
\vspace*{0cm} {\it \begin{center}
    Theoretical Physics Group \\
    Ernest Orlando Lawrence Berkeley National Laboratory \\
    University of California, Berkeley, California 94720, USA
    \end{center} }
\vspace*{1cm}

\end{center}

{\Large \bf \begin{center} Abstract \end{center} }

We reconsider the options for cosmological holography. We suggest that
a global and time--symmetric version of the Fischler-Susskind bound is
the most natural generalization of the holographic bound 
encountered in AdS and De Sitter space.
A consistent discussion of cosmological holography seems to imply an
understanding of the notion of ``number of degrees of freedom'' 
that deviates from
its simple definition as the entropy of the current state. The introduction
of a more adequate notion of degree of freedom makes the suggested
variation of the Fischler-Susskind bound look like a stringent and viable bound
in all 4--dimensional cosmologies without a cosmological constant.

\renewcommand{\baselinestretch}{1.2}

\newpage
\renewcommand{\thefootnote}{\arabic{footnote}}
\setcounter{footnote}{0}


\section{Introduction}
\label{intro}

For quite some time the remarkable fact that the entropy of a black hole (BH)
is represented by its event horizon \cite {bh} has nurtured the suspicion 
that gravity in general could be holographic \cite{th-s}, meaning that all 
degrees of freedom (DOF) of a gravitational theory 
can be projected onto some lower dimensional surface(s) whose storage 
capacity is one bit information per Planck area. 
The discovery of AdS/CFT duality \cite{m-g-w,agmoo} strengthened this 
suspicion by directly
identifying a gravity theory on AdS with a lower dimensional field theory
on its boundary.

In \cite{fs} a specific realization of a general cosmological holographic 
principle was suggested. A very attractive point of that proposal 
is the fact
that the suggested holographic bound is exactly saturated at the Planck time
if one straightly extrapolates back from today's condition of the universe. 
Less satisfying
seemed the fact that today the actual entropy inside the particle horizon
is far lower than the FS-bound. The most serious problem 
however was the fact that in a closed universe the FS-bound seemed necessarily
violated. During the last year various different modifications of the 
FS-conjecture have been formulated to deal with the described problems 
\cite{HB}. 

In this paper we reconsider the options for a cosmological
holographic principle.
In the first part we investigate what implications for a cosmological 
holographic principle one can infer from the realization of holography
in AdS (section 2) and De Sitter space (section 3). We claim that the most 
natural generalization would be a principle that it is global, in the
sense that it does not restrict the local density of DOF, and time symmetric. 

Section 4 makes the statement, following the basic idea of \cite{spec}, 
that a consistent discussion of cosmological holography requires a
reconsideration of the notion of degree of freedom (DOF) in that context. 
The recent discussions of a cosmological
holographic principle have been based on the assumption that the actual
entropy of a system represents the number of degrees of freedom that
have to be stored on some boundary in a holographic scenario.
This is true as long as one deals with field theories in
an equilibrium state disregarding gravity - in which case there is no
holography anyway. However we argue that the
assumption becomes quite questionable in a general cosmology. There are two
aspects that change the situation. First a realistic cosmology
is not in the highest entropy state because matter is more or less evenly 
distributed and gravitational clumping can enhance the entropy.
It is not a priori clear whether the possible DOF must be
identified with the actual entropy or the highest possible entropy.
Second, the background as well as -- if defined in some dynamical way --
the position of the holographic boundary depend on the dynamics happening
on that background. Therefore the validity of the background and the 
position of the boundary both can give
restrictive conditions for the number of DOF. The central
statement of this paper is that a definition of ``degree of freedom''
that takes into account the two aspects mentioned above seems to
render a global and time--symmetric version of the original Fischler-Susskind 
conjecture viable and rather stringent for all four dimensional
cosmologies with zero cosmological constant.

Section 5 describes the
problems one encounters in general cosmological scenarios 
with a cosmological
constant. We close with conclusions.

\section{Cosmological Holography}


The basic claim of a cosmological holographic principle is
that some kind of holographic bound generally exists in cosmology.  
The upper bound for the number of degrees of freedom
of a three--dimensional space--like or light--like space should be 
represented by the area of some two dimensional surface on which one bit 
information per Planck area can be stored.

The idea was first formulated in the early papers of t' Hooft and Susskind
\cite{th-s} and was more specifically realized by Fischler and Susskind
in \cite{fs} (FS--conjecture). The conjecture goes as follows:
Given some spherical spatial volume $V$ bounded by the surface $B$,
all DOF entering through the future--lightcone that ends at $B$ can be stored
on $B$ if $B$ can store one bit information per Planck area.

There exists one rather powerful argument in favour of the
existence of a cosmological holographic principle along those lines:
The corresponding holographic bound turns out to be valid in 
the past of our universe
up to the Planck time and moreover seems to be more or less exactly fulfilled
at the Planck time itself. This is a quite remarkable coincidence.

However on the other side the FS--bound has some problems.

First it seems that the bound is violated for a 
boundary $B$ around the center of a BH whose radius is smaller 
than the BH--event horizon.  
Second it seems not very satisfying
that today the actual entropy inside the particle horizon
is far lower than the FS-bound. 
The most serious problem 
however is the fact that in a realistic closed universe the FS-bound 
seems necessarily
violated. This has to do with the fact that in a matter dominated
closed 4-dimensional
Robertson Walker spacetime light starting from one point at $t=0$ reaches 
the opposite pole on the sphere at the time of maximal expansion.
Therefore already at times close to the maximal expansion there exist 
arbitrarily
small $B$s around the opposite pole which have to store nearly all of the 
universe's DOF, which seems impossible.

During the last year various different modifications of the 
FS-conjecture have been formulated to deal with the described problems 
\cite{HB,el,b}. The most recent interesting attempt to construct a
holographic principle that is generally valid was made in \cite{b}.
However the concept has some essential problems in connection with
black holes which are partly discussed
in \cite{l}. In particular \cite{l} found an explicit example where
the bound of \cite{b} does not hold.

As the situation stands today, it seems to us that no cosmological 
holographic principle is enforced by known gravity.\footnote{What seems to
be true in all gravitational scenarios as long as one accepts the viability
of statistics is the generalized second law formulated in \cite{el}.
This however is no holographic principle.)}

However a holographic principle might exist without being implied by 
semi-classical gravity. The one known example of a full holographic theory,
AdS/CFT duality, was conjectured in the context of D--branes and 
does not follow from classical gravity.
The suspicious frequency of holographic phenomena in connection with
gravity really looks like a hint towards a general 
holographic principle that can only be understood in
full quantum gravity. In that case this principle could well put new
constraints on the cosmological evolution that do not exist in classical
gravity. Without knowledge of quantum gravity the only way to understand
in how far such a possibility is realistic is to check whether 
our universe as we know it is consistent with some holographic principle 
and possibly shows apparently unnatural constraints that can be understood
in accordance with holography.
This as we understand it is the spirit of the original work of 
Fischler and Susskind. 
In this paper we try to go a little further in that direction. 

A good point of departure seems to be the known example
of AdS/CFT duality. We look for a definition of cosmological holography 
that remains as close as possible to this example. 
If we compare the FS--conjecture or Bousso's covariant entropy
conjecture with the situation
in the case of AdS/CFT duality, we realize 
that the former are much stronger in a certain point. 
Since they allow for all possible $B's$,
also very small ones, they give a bound on the local density of DOF,
something AdS/CFT duality with the boundary strictly at infinity
does not do. And it is exactly this requirement that produces considerable
problems in connection with black holes. Additionally this requirement 
devaluates the strongest cosmological argument in favour of the
FS--bound, namely the fact that the FS--bound is exactly saturated at the 
Planck time:
If it was the case that the FS--bound for some reason holds even
for the smallest boundary $B$ and the highest energy density anywhere in 
space, the statement
that already the ``naive'' estimate in \cite{fs} leads to a saturated
bound at the Planck time could only be judged as a mere accident.

The arguments above lead us to the assumption that a 
cosmological holographic principle, if it exists, 
is likely to be a global principle in the sense that it does not give
bounds on the local entropy density.
Local realizations of holography can occur as a consequence of the
global principle but do not necessarily hold under all circumstances.

To make the FS--bound global one just has to require that the surface $B$ must
be the particle horizon respectively past event 
horizon of some observer. This definition makes the holographic conjecture
immune against local problems with today's black holes. $B$ is small
only close to the Planck time.\footnote{In inflationary scenarios 
the observed initial state is the state after re-heating. In this case 
one would hope to find an argument to replace the exponentially blown up 
and therefore quite meaningless particle horizon with an ``effective''
particle horizon reaching back to the endpoint of the exponential expansion.}
The holographic conjecture therefore enforces the specific energy density 
of the universe one finds at 
the Planck time. It also enforces that the universe is
smooth close to the Planck time avoiding high entropy states that include
primordial black holes. Holography in this form therefore implies 
cosmic censorship.  

\section{Pure De Sitter Space}


A main inspiration for the formulation of the holographic bound was
AdS/CFT duality. However there is a fundamental difference between
the static AdS scenario and a realistic cosmology:  
The AdS structure is only valid as long as the global energy density
is zero. Otherwise the Einstein equations would enforce a different spacetime
structure. Thus it  is conceptually inconsistent to try to project states
with finite global energy density onto the AdS boundary.\footnote{It seems
to us that the connection between AdS/CFT and this condition has its
subtleties. A discussion of that point is in preparation.}
Seen from that perspective the step to cosmological holography is the
step to finite matter density.

Besides AdS (and Minkowski space that can be understood
as it's infinite radius limit) there exists one more example for a zero
matter density cosmology: De Sitter space. In this section we will discuss
whether the principle of holography encountered in AdS and generalized to
Minkowski space can also be implemented in De Sitter space.
From this discussion we will get further hints concerning a consistent
possible generalization of holography in a realistic cosmological context.

The spacelike dimensions of De Sitter space are closed which 
has some implications for the 
character of a holographic bound in this scenario. 
Since the space volume is finite, the condition of background validity
requires empty space. Any small matter contribution would change 
the background. While in the AdS case the holographic boundary of AdS sat at 
infinity, there is
no infinity in de Sitter space. Therefore a holographic boundary will have to
sit somewhere in space, thereby resembling already the situation we will
encounter in finite matter density scenarios.

Let's have a closer look at the structure of De Sitter space. 
De Sitter space is the solution of the state equations

\bea
\frac{\Lambda}{3}R^2-1 &=& \dot{R}^2 \label{dsse} \\
\Lambda R &=& 3 \ddot{R}
\eea

Its metric can be written as

\beq
ds^2=-dt^2+\alpha^2 cosh^2(\alpha^{-1}t)d\Omega , ~~~
\alpha :=(\frac{3}{\Lambda})^{1/2}
\label{dsm}
\eeq

Eq.(\ref{dsse}) leads to the Hubble constant

\beq
H :=\frac{\dot{R}}{R}=(\frac{\Lambda}{3}-R^{-2})^{1/2}
\label{dshc}
\eeq

De Sitter space has a future and a past event horizon $E_f$ respectively
$E_p$. They are defined as

\bea
E_f(t) &:=&R(t)\int_{t}^{\infty} \frac{dt^\prime}{R(t^\prime)} \\
E_p(t) &:=&\int_{-\infty}^{t} \frac{dt^\prime}{R(t^\prime)}
\label{eh}
\eea

Solving Eq.(\ref{dsse}) gives

\beq
R(t) = \frac{1}{H} e^{\left(Ht-\frac{ln4}{2}\right) }+
\frac{1}{4H}e^{-\left(Ht-\frac{ln4}{2}\right) }
\label{r}
\eeq

which leads to 

\bea
E_{f(p)}(t)/R(t)=\left. \frac{2}{H}\arctan\left[2  
e^ {\left(Ht-\frac{\ln4}{2}\right)} 
\right]  \right|_{t(-\infty)}^{\infty(t)} 
\label{ehe}
\eea

This implies 

\bea
E_f(\infty)=E_p(-\infty)&=& R(0)=H^{-1} \\
E_f(0)=E_p(0)&=&\frac{\pi}{2H} 
\label{ehee}
\eea

Therefore we have the following situation: At $t=-\infty$ $E_p$ is at
a distance $H^{-1}$ from the observer while $E_f$ is at the
same distance from the opposite pole. They meet each other at $t=0$ at the
equator and end up with exchanged roles at $t=\infty$. Each horizon travels
a coordinate distance $\pi$ during its infinite lifetime.
This situation is directly related to the AdS case. The corresponding statement
there would be that light reaches infinity at $t=\pi/2$.

It was shown in \cite{gh2} that in the exponentially expanding phase
of De Sitter space the area of $E_f$ plays the role 
of entropy exactly in the same way as in the case of the black hole.
Since De Sitter space is empty, it would be natural to have a holographic 
bound that is exactly saturated by the geometric entropy. 
The holographic bound should resemble the
area of the event horizon. Now the future event horizon is supposed to
cover the DOF outside the horizon seen from the observers perspective.
It therefore fits exactly as holographic bound around the point 
opposite to the observer's position on the sphere. The discussion above
shows us that an area that resembles the future event horizon for one point
represents the past event horizon $E_p$ for the point opposite on the sphere.
Consequently $E_p$ is a natural candidate for the holographic bound on De
Sitter space in the expanding phase.\footnote{Identical to the past event
horizon in De Sitter space would be
the outer apparent horizon that separates the anti-trapped region from the
normal region around the other pole. This outer apparent horizon
does not exist in open universes and therefore cannot serve as a general
definition of a holographic boundary. It can however play a role in a
definition of a holographic boundary together with the particle horizon. See
\cite{b}.} \footnote{Like its cousins with negative respectively 
zero cosmological constant, the holographic boundary of De Sitter space has a
constant area.}

Now, interestingly, in the contracting phase the whole situation is reversed.
By changing exponential expansion to exponential contraction one changes 
the area $r=1/H$ from representing $E_f$ to representing $E_p$.
Since the definition of entropy does not depend on the time direction,
now the De Sitter entropy
must be defined by $E_p$. But the observer's $E_p$ is the opposite point's
$E_f$. Therefore the holographic bound is represented by $E_f$. 
We get the remarkable result that the definition of the holographic boundary
depends on the sign of the expansion of space. $E_p$ and $E_f$ change roles
at the point of extremal extension. This point will play an important role
in the discussion of cosmological holography.

In summary De Sitter space has a very natural holographic boundary
that is consistent and stringent and resembles the global FS-bound in a
time symmetric form. The closed structure of space
in itself is not contradictory to holography.

\section{What is a Degree of Freedom}


We know that the scenarios with zero energy density are holographic.
The way holography is realized there led us to the following concept
for a possible general holographic principle:

Assume $B$ to be an observer's past horizon in the expanding phase
of the universe respectively an observer's future horizon in the contracting 
phase of the universe. Then all DOF which entered through the future light cone
respectively will exit through the past light cone ending on $B$
can be stored on $B$ if $B$ can store one bit information per Planck area.

Next we have to check whether this holographic principle can hold in our
real world. The crucial point will be how to deal with the closed universe
scenario.

One question is of central importance in order to know what we are doing:
We must have a clear understanding of what is a DOF. This turns out to be
less trivial than it seems.

We saw already in De Sitter space that the background validity condition 
enforces empty space, which in turn is reflected
by the fact that the holographic boundary is exactly saturated by geometric
entropy. This is a first example showing that the validity of a holographic 
bound can be intricately entangled with cosmological restrictions.

In a realistic cosmology the situation is even more subtle. The evolution
of space as well as the evolution of a holographic boundary in terms of 
comoving coordinates depends on the matter and radiation energy in that space. 
But to make the statement what
DOF can be stored on a given surface, it is necessary first to know
where this surface is. Holography can only talk about 
DOF under the precondition that the boundary is where it is.
Therefore the full history of the universe leading up to the current
position of $B$
has to be considered in deciding which DOF are to be stored on $B$. 

In order to deal with this complication we introduce the notion of
``relevant degree of freedom'':
Assume some cosmology and some holographic boundary surface 
put on the spacetime of that cosmology. Those states of the system 
which are consistent with the given evolution of
both the universe and the holographic boundary are called relevant DOF
for that boundary. Only relevant DOF are to be stored on a holographic
boundary.

Note that the relevant DOF for a boundary are still to be distinguished from 
those which can be stored on that boundary.
The question which subset of the relevant DOF eventually are accounted
for by the holographic boundary depends on the specific definition
of the holographic conjecture. 

How exactly to define ``consistent with the evolution'' 
in the definition above is not quite 
clear. The closer one looks, the less tractable
situation becomes. The dynamics of the holographic
boundary  depends not just
on the global features of spacetime but also to some extent on the
local conditions at a given point. The problem is 
reminiscent of those essential problems in gravity connected to the fact 
that what one would like to use as background space is actually part of 
the dynamics. 
The only way to deal with this situation at the moment seems to be to
neglect small local effects and just look for intuitively plausible
global restrictions.

We consider a
flat spacetime evolving from a big bang according to Einstein's equations
and look at the situation at some time $t$. What are the relevant
DOF of this system? The global structure of spacetime is defined by the 
initial
parameters of the cosmology. Higher energy density would imply a closed
space, lower energy density an open space. Both would imply a 
different size and position of $B$ which means that both
cannot represent relevant DOF of any $B$ in flat space. 

The position of $B$ won't be influenced by the degrees of freedom which
represent the entropy of the state (at least to a good approximation). 
Therefore those must be counted as 
relevant DOF. The crucial question is the following: 
Is the number of relevant DOF inside a surface $B$ represented by
the actual entropy of the system or by the maximal entropy state possible.
Before dealing with the question we have to make an assumption about
the maximal entropy state of the system. Given a certain energy content,
the highest possible entropy will be connected to the largest possible black
hole. Since holography enforces an initial state without black holes,
all black holes must form dynamically and thus cannot be 
arbitrarily large. We assume that the maximal entropy state corresponds
to a black hole with the mass 

\beq
M_{max}(t)\sim \rho (t)H^{-3} \sim \rho (t)E_p^3
\eeq

where $\rho(t)$ is the global energy density. 
The maximal entropy inside $B$ therefore should be represented by a BH that
contains all energy inside $B$. The example of AdS suggests that the DOF 
of this system should be stored at the boundary. 
However the situation requires closer examination: 
Holography forces the initial state
to be a smooth state that has no black holes and whose radiative entropy 
is in equilibrium.
If one treats the system classically, all evolution is strictly
deterministic, therefore the number of DOF in a system has to remain
constant. At a later stage after black holes have formed this would
mean that a big number of BH states would be locally allowed but
do not stem from a purely radiative initial state. Thus they
are not relevant DOF. This argument would reduce the number of relevant DOF
to the value of radiative entropy. 

Quantum processes however
do produce new DOF. An extreme example is inflation where the quantum
process of re-heating produces basically all DOF of today's world.
Assume there exists an initial microstate at the Planck time
that leads to a single black hole storing all matter inside $B$ at time 
$t$.  One would 
expect, without making any more specific statement about BH-physics, 
that eventually all BH states can be reached through quantum tunneling
in the BH. Therefore one should conclude that all BH DOF are relevant.
This does not yet mean that they are all to be accounted for at the boundary.
The FS--condition says that only those DOF are to be stored which enter through
the horizon. If a BH is develops inside the horizon and new BH DOF are
produced, these DOF cannot be stored at the horizon. However
every BH at any moment can be thought of as entering some particle horizon
around an observer somewhere else.\footnote{Note that this is fundamentally 
different
from the case of DOF production through re-heating. The latter is an isotropic
process that looks the same for any particle horizon. Consequently the
newly produced DOF inside the blown up particle horizon are truly 
``invisible'' to the holographic boundary.} 
We conclude that the holographic bound has to cover a number of DOF whose
order of magnitude is somewhere close to the entropy of the maximal BH. 

How does this apply to the actual universe?
A rough estimate says that the maximal BH entropy inside the particle horizon
today is

\beq
S_{max}(today) \sim 10^{123}
\eeq

while the area of the horizon in Planck units is

\beq
A_H(today) \sim t(today)^2 \sim 10^{122}
\eeq

The bound looks realistic and plausibly saturated. The next question is in
how far this remains the case in the past respectively the future.

The radius of the particle horizon moves like 

\beq
r_{PH}(t)\sim t
\eeq

In the matter dominated regime the line element grows in $d$ spatial 
dimensions like

\beq
{R}(t) \sim t^{\frac{2}{d}}
\eeq

and we have a constant energy per comoving volume

\beq
\rho (t) R^d(t) = K
\eeq

This means that the energy inside the horizon grows like

\beq
M_{PH} \sim \rho (t) r_{PH}^d(t) \sim Kt^{d-2}
\eeq

On the other hand the BH mass goes like

\beq
M_{BH} \sim r_{BH} 
\eeq

and the BH entropy goes like

\beq
S_{BH} \sim r_{BH}^{d-1}
\eeq

If we identify the maximal entropy $S_{max}$ with the black hole built
from all energy inside the horizon, we see that 

\beq
\frac{S_{max}(t)}{A(t)} \sim t^{(d-3)(d-1)} , ~~~~for~~~ t>t_D
\eeq

where $t_D$ is the decoupling time from which on the dynamics is matter 
dominated. For $d=3$ one gets

\beq
\frac{S_{max}^{d=3}(t)}{A^{d=3}(t)} = const , ~~~~for~~~ t>t_D
\eeq

This means that if saturated today, the holographic bound
remains saturated for all time into the future and back until $t_d$.

One has to note that this nice result is only true for 3
spatial dimensions. In particular $\frac{S_{max}(t)}{A(t)}$
increases with a positive power of $t$ for $d>3$ and thus necessarily leads to
a violation of the holographic bound at some time.

The case of radiation domination for $d=3$ with 

\beq
{R}(t) \sim t^{\frac{1}{2}}
\eeq

leads to

\beq
\frac{S_{max}(t)}{A(t)} \sim  t , ~~~~for t<t_d
\eeq

Therefore $\frac{S_{max}(t)}{A(t)}$
would decrease if one goes further into the past beyond $t_d$. 
However in that regime it clearly makes no sense to 
consider BH entropy as relevant DOF since radiation domination is
conceptually exactly opposite to BH domination. 

The case of an open universe initially cannot be distinguished from the
flat case. At late stages the ratio $\frac{S_{max}(t)}{A(t)}$ will 
become considerably smaller than in the flat case. Assuming that the
holographic bound is saturated in the flat case, in the open case the
bound will be nearly saturated at the early stages of matter domination
while the number of DOF will fall below the saturation level
later.

The most interesting scenario is the closed universe. This is where a
holographic bound is most difficult to define. In \cite{fs} where 
the degrees of freedom were identified with the actual entropy of the
universe at a given state, it was found that a holographic bound 
defined by the particle horizon necessarily is violated at late stages
when the boundary has entered the opposite half of the spatial sphere, 
therefore
decreasing its coordinate area until, when it reaches the opposite point, it
eventually becomes zero. In the discussion of the flat case we argued that
the current entropy underestimates the relevant DOF. Now we will argue
that in the closed case it is a radical over--estimation. 

At this point we have to remember the time symmetric form of our 
holographic conjecture. 
The definition of the holographic bound is connected to the sign of
the universe expansion. The holographic bound for the 
contracting phase is given by the future event horizon. We have
seen that holography enforces a smooth initial state without initial 
black or white holes. Now this implies that we also must have a smooth 
final state. But this means that existing black or white holes have to
dissolve in the contracting phase. The entropy arrow must be 
reversed there. Holography connects the entropy arrow to the
sign of the universe expansion. 

What does this mean for the DOF? Let us consider a certain macrostate
at the time of maximal expansion ($t_{max}$). 
The microstates of this macrostate represent the entropy at that stage. 
Now due to the time symmetric nature of holography,
relevant DOF are only those which are consistent with the universe evolution
seen from both time directions. In other words the requirement that the system
has to return to a lower entropy state at the final state
has to be fulfilled by each state that represents a relevant DOF.
This however will rule out most of the microstates at time $t=t_{max}$.  
Only the small subset that leads back to the initial condition represents 
relevant DOF.\footnote{It is highly nontrivial
to formulate this statement more exactly in the framework of quantum
statistics (See some considerations in \cite{spec}). In this paper we will 
not go deeper into that matter.}
The consequences are clear. It is not justified to use the entropy as
a measure for relevant DOF in a closed universe. Let us see how this fits
into the cosmological picture.

The evolution of a closed (d+1)--dimensional spacetime goes like

\bea
R(\chi) \sim (sin|\kappa \chi |)^{\kappa^{-1}},~~~~~~~
\kappa=\frac{d}{2}(1+\gamma)-1
\eea

where $\chi$ is the coordinate 
position of the particle horizon,

\bea
\chi(t):= \int_0^t \frac{dt^{\prime}}{R(t^{\prime})}
\eea

and $\gamma$ is the ratio between pressure and matter density.
In 3 spatial dimensions we therefore have the following situation:
In the matter dominated case with zero pressure we have $\gamma=0$
and $\kappa=1/2$. At the time of maximal expansion, characterized by
$\kappa\chi =\pi /2$ we have $\chi= \pi$.
The particle horizon reaches the opposite pole 
at the time of maximal expansion.
In the contracting phase the future event horizon plays the role of
the holographic
bound making the situation time symmetric. Since holography enforces
an initial radiation dominated state in fact the holographic boundary
will not quite reach the pole, meaning that at the time of maximal
extension there still remains a small area for the boundary. This 
fits nicely into the picture described above.

In a scenario where the universe remains radiation dominated throughout its
evolution, we have $\gamma=1/3$
so that $\kappa=1$ and $\kappa\chi = \chi$.  The horizon just reaches the 
equator at the point of maximal expansion.
Therefore the bound at the time of maximal expansion is not much tighter
than it would be at the same time in a flat scenario. If the initial
state is purely radiative, the radiative entropy can be stored at
the boundary throughout the whole evolution of the universe.
 
If we understand radiation domination in its extreme realization, 
we have have a situation where gravitational clumping does not happen at all. 
There is no gravitational
entropy. The radiative (as the much lower baryonic) entropy however is in 
an equilibrium
state from the start, therefore the initial and final conditions do not
pose any restrictions. This means that the radiative entropy in this case
in fact represents the relevant DOF, which again nicely fits into
the picture.

One has to notice that this nice situation once again only holds for $d=4$.
The defined holographic principle seems inconsistent with lower dimensional 
cosmologies while for higher dimensional cosmologies the holographic bound
evolves less far on the sphere and therefore does not fully reflect the 
constraints due to recontraction. 

\section{Cosmological Constant plus finite Energy Density}


In this chapter we have a short look at the most general case of cosmologies
where a cosmological constant $\Lambda$ is joined by finite energy density.
In that case the statement that the universe is closed 
and the statement that it recontracts do not
imply each other. Therefore the nice connection between the shrinking
of the the holographic boundary around the other pole and the reduction
of relevant DOF due to recontraction does not work any more. 

A cosmology with negative $\Lambda$ always recontracts.
Therefore one would expect a reduction of relevant DOF similar to
the closed cosmology with $\Lambda =0$. However cosmologies with
negative $\Lambda$ can be flat or open so that the boundary does
not shrink. It seems that the holographic bound is unreasonably high. 

The situation in case of positive $\Lambda$ is even worse.
The problematic case is the closed scenario.
Let us assume a matter content sufficient to make the universe recontract.
If we ad a small  $\Lambda$, the effect will be to drag out the large expansion
phase before recontraction takes over. This means that the particle horizon
will - in the matter dominated case - reach the opposite pole earlier than 
the time of maximal expansion and vanishes. This would imply that there are no 
DOF at all. This is most difficult to understand if $\Lambda$ is
slightly above the value necessary to prevent recontraction. In that case
there will be a phase of semi-stability at a certain expansion followed
by the transition to exponential expansion. The particle horizon will     
hit the other pole in the semi-stable phase even though the restrictions
on relevant DOF stemming from recontraction do not apply.
For even higher $\Lambda$ we have the case of a bouncing universe
looking similar to pure De Sitter space. In that case the horizons that
should represent the holographic bound also vanish at some time 
at the point on the sphere opposite to the observer.

\section{Conclusions}


We have formulated what we believe to be the most straightforward 
generalization of the holographic bound one encounters in cosmologies with
zero matter density. We further argued that the DOF relevant for holography
are not simply the ones represented by the actual entropy.
The resulting picture of a global time--symmetric holographic principle 
and a set
of relevant DOF does not fit into most cosmological scenarios with finite 
energy density. The holographic bound fails for a flat universe in more
than 4 spacetime dimensions, for a closed
universe in less than 4 spacetime dimensions 
and for a closed scenario in 4 dimensions
with a considerable positive cosmological constant. On the other side the bound
it imposes is far higher than necessary for most of the other scenarios.

However there is one scenario where the holographic bound is both viable
and stringent. This scenario is the 4--dimensional universe with a strongly 
suppressed cosmological constant we actually live in. It is surely too early
to judge whether this is just a coincidence or not. If it is more than a
coincidence, it would mean that holography hints towards an 
excitingly strong restrictiveness of full quantum gravity.


{\bf Acknowledgments:} We would like to thank Paolo Aschieri for very
useful discussions. This work was supported in part by the Director,
Office of Science, Office of High Energy and Nuclear Physics, 
Division of High Energy Physics,, of the 
U.S. Department of Energy under Contract DE-AC03-76SF00098 and in part
by the Erwin Schr\"odinger Stipendium Nr. J1520-PHY.

\newpage


\parskip=0ex plus 1ex minus 1ex


\end{document}